\documentclass[slac_one]{revtex4}
\usepackage{graphicx}
\usepackage{fancyhdr}
\begin{document}

%Title of paper
\title{The Time Projection Chamber for the ALICE Experiment}

\author{C. Lippmann}
\author{the ALICE TPC Collaboration}
\affiliation{CERN, 1211 Geneva 23, Switzerland}

\begin{abstract}
 
  The Time Projection Chamber of the ALICE Experiment has been installed in
  the experimental setup in the underground area at the Large Hadron Collider at
  CERN in Geneva. The Alice TPC ReadOut (ALTRO) chip implements intelligent signal
  processing on the Front-End-Electronics. During the years of 2007 and 2008
  commissioning and calibration of the TPC have been carried out with cosmic
  rays, radioactive Krypton isotopes and with tracks produced by a UV laser system.
  In addition to these in this publication we present first results on energy loss
  measurements and on the momentum resolution.

\end{abstract}

\maketitle

\thispagestyle{fancy}

\section{INTRODUCTION}

ALICE\,\cite{alice} will search for evidence of the quark-gluon plasma---a state
of matter which is believed to have existed just after the Big Bang---in head-on
collisions of lead-ions at the LHC. It will also be used to study the properties of
proton-proton collisions. The trajectories of thousands of charged particles produced
in central lead-lead collisions have to be measured and the particles have
to be identified. To serve these tasks, the largest Time Projection Chamber (TPC)
in the world was installed in the central barrel of ALICE\,\cite{tpctdr}.

The ALICE TPC provides track finding, momentum measurement and PID
at transverse momenta 0.1\,$<$\,$p_t$\,$<$\,100\,GeV/c and in the pseudorapidity
range $|\eta| < 0.9$.
The requirements in the extreme multiplicities of ion collisions at the
LHC are however very challenging: A good tracking efficiency ($> 90$\,\%) and
a momentum resolution of $\approx$1\,\% at 2\,GeV/c (TPC only),
$\approx$10\,\% at 50\,GeV/c (TPC only) or $\approx$3.5\,\% at 100\,GeV/c
(TPC together with the inner tracking system and Transition Radiation Detector TRD).
The d$E$/d$x$ resolution has to be better than $10$\,\% and the rate capability
has to be $200$\,Hz for central lead-lead collsions and $1$\,kHz for proton-proton
collsions.

The TPC is cylindrical in shape with an active radial range from 85 to 247\,cm and an
overall length of 500\,cm. A central high voltage electrode divides the volume into
two drift regions of 250\,cm length. As readout chambers (ROCs) multiwire proportional
chambers with cathode pad readout are mounted in 18 trapezoidal sectors in both
endplates. Each sector is segmented in two ROCs (Inner and outer - IROC and OROC) with
a pad (and wire) geometry optimized to the expected occupancy (below 40\% at the maximum
design multiplicity of d$N_{ch}$/d$\eta$=8000). In total the TPC has 557\,568 pads with
three sizes; the smallest pads are $4\times 7.5$\,mm$^2$. A light cold drift gas
(Ne, CO2, N2) was chosen because of the small diffusion and low multiple
scattering\,\cite{chilo}. It also has a low number of ionisation electrons per
unit length, which together with the rather high ion mobility helps to minimise field
distortions due to space charge for running at high luminocities with high rates of
secondary charged particles. However, the low ionisation energy loss in combination with
the chosen pad sizes requires a rather large gas gain of up to 10$^4$ in order to
meet the requirement of a good $S/N$ ratio of $\geq 20$.

The construction and assembly of the TPC was completed in 2006. After
an extensive (pre)commissioning phase on the surface the TPC was lowered into
the underground experimental area and transferred into the large solenoidal
magnet in January 2007, where commissioning of all service infrastructure and
readout electronics took place. Since then the TPC was calibrated with laser
and cosmics tracks and with radioactive Krypton decays in the active gas volume.
The different calibration procedures are described in Section \ref{seccalib}.

The task of large acceptance tracking in the ALICE TPC is similar to that
encountered in previous heavy-ion experiments at the SPS and RHIC. However,
the extreme multiplicities of ion collisions at the LHC set qualitatively
and quantitatively new demands. This made indispensable the design of a
new read out chip, which will be described briefly in the following section. 

\section{ALICE TPC READOUT CHIP}

\begin{figure*}[t]
\begin{minipage}[h]{8cm}
\includegraphics[height=7cm]{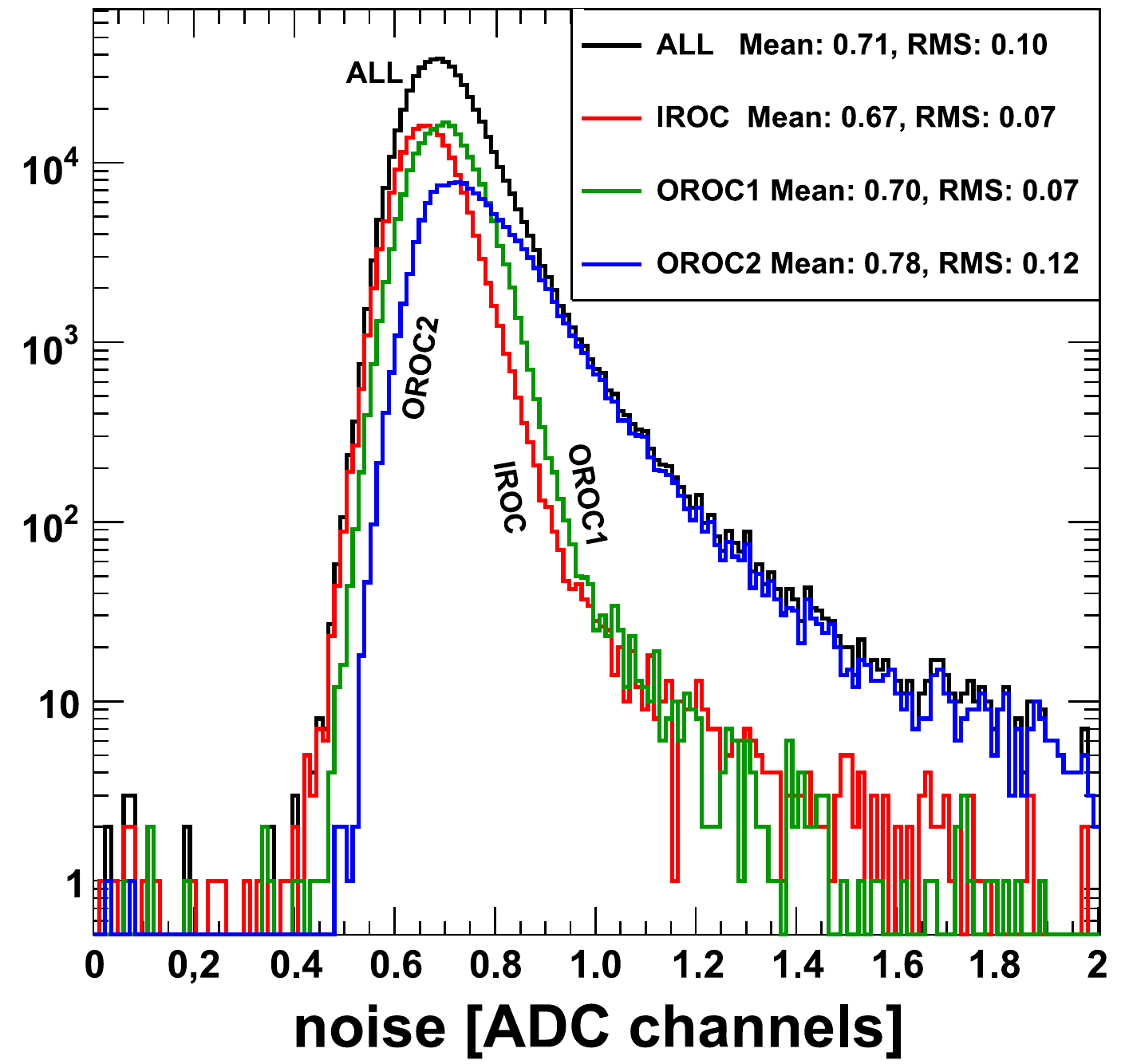}
\caption{The distribution of the overall system noise in the final setup in
  the ALICE experiment. Histograms are shown for the three different pad sizes.
  OROC1 and OROC2 indicate the two regions of the OROCs with medium and large
  pad sizes.}
\label{noisepic}
\end{minipage}
\hfill
\begin{minipage}[h]{8cm}
\includegraphics[height=7cm]{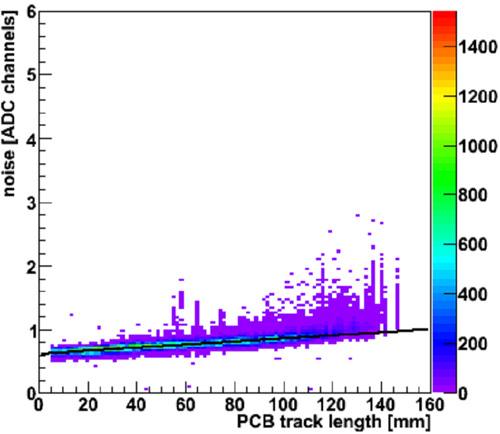}
\caption{Correlation of the system noise with the length of the traces
  on the pad plane PCB. The noise is clearly correlated to the capacitances of the
  individual read out channels, which is proportional to the length of the readout
  traces (PCB track length). The solid line is a fit to the data.}
\label{corrpic}
\end{minipage}
\end{figure*}

The Alice TPC ReadOut (ALTRO) chip\,\cite{musa,altro} implements a 10 bit, 10\,MSPS
ADC and digital filtering circuits in four stages for 16 channels. The first
stage implements a baseline correction in order to prepare the signal for the
tail cancellation; low frequency perturbations and systematic effects are removed.
The tail cancellation block is used to suppress the tails of the pulses within
1\,$\mu$s after the peak with very good accuracy. The filter coefficients for each
channel are fully programmable and
thus the circuit is able to cancel a wide range of signal tail shapes. The third
processing block applies a baseline correction scheme based on a moving average
filter, removing non-systematic perturbations of the baseline that are
superimposed on the signal. At the output of this block, the signal baseline
is constant with an accuracy of 1\,ADC count, which makes possible a very efficient
zero-suppression. This is important because it largely reduces the data volume
while keeping the interesting signals. The readout is finally performed out of
the on-detector event buffers upon receipt of a level 2 trigger accept signal.

\section{TPC CALIBRATION} \label{seccalib}

After the comissioning of the overall system is finished important tasks are
calibration and alignment. We use data that are aquired in dedicated calibration
runs. Some of the calibration methods used are described in the following.

\subsection{PEDESTALS AND NOISE}

Pedestal runs are taken periodically. A histogram of the overall system noise
in all TPC readout channels is shown in Fig. \ref{noisepic}. Noise is defined
as r.m.s. of the baseline. The present value is clearly better than the design
value of 1 ADC count, the mean noise is about 0.7 ADC counts and as Fig.
\ref{corrpic} shows the noise is correlated to the capacitances of the individual
readout traces. This shows that the system noise close to the natural limit. The
pedestal values are very constant with time. 

\subsection{ISOCHRONICITY}

Small variations in the signal arrival time ($t_0$) are due to chip-to-chip
variations of the signal shaping and due to signal delays. They can be studied
and calibrated using pulses injected into the cathode wire grid of the ROCs.
The induced signals are read out from all pads and the time position and
amplitude of the pulser signals are analysed event-by-event.

\begin{figure*}[t]
\begin{minipage}[h]{8cm}
\includegraphics[height=7cm]{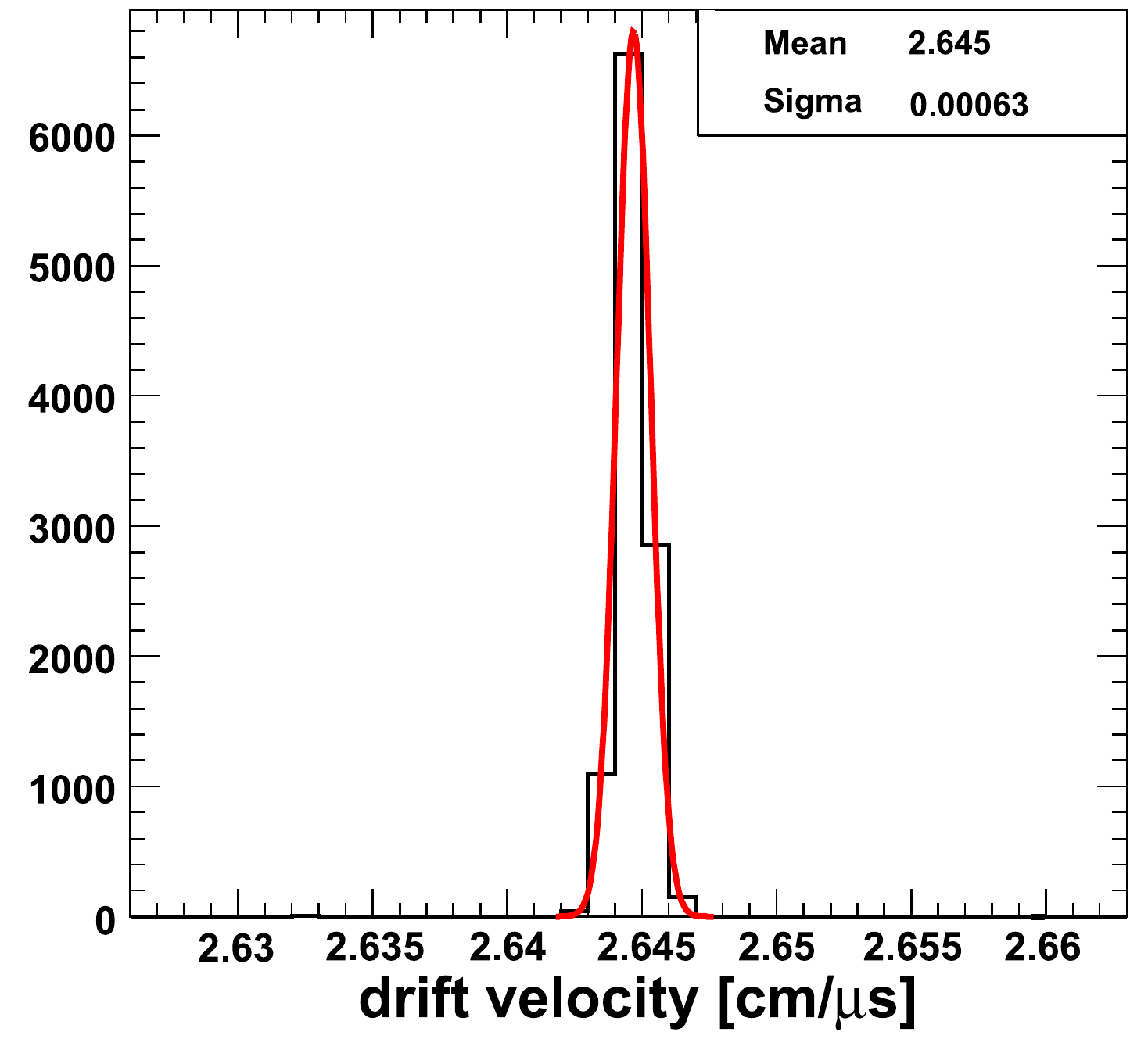}
\caption{Mean drift velocity as calculated from the central electrode signals.
  The relative resolution is $2\cdot 10^{-4}$.}
\label{vdpic}
\end{minipage}
\hfill
\begin{minipage}[h]{8cm}
\includegraphics[height=7cm]{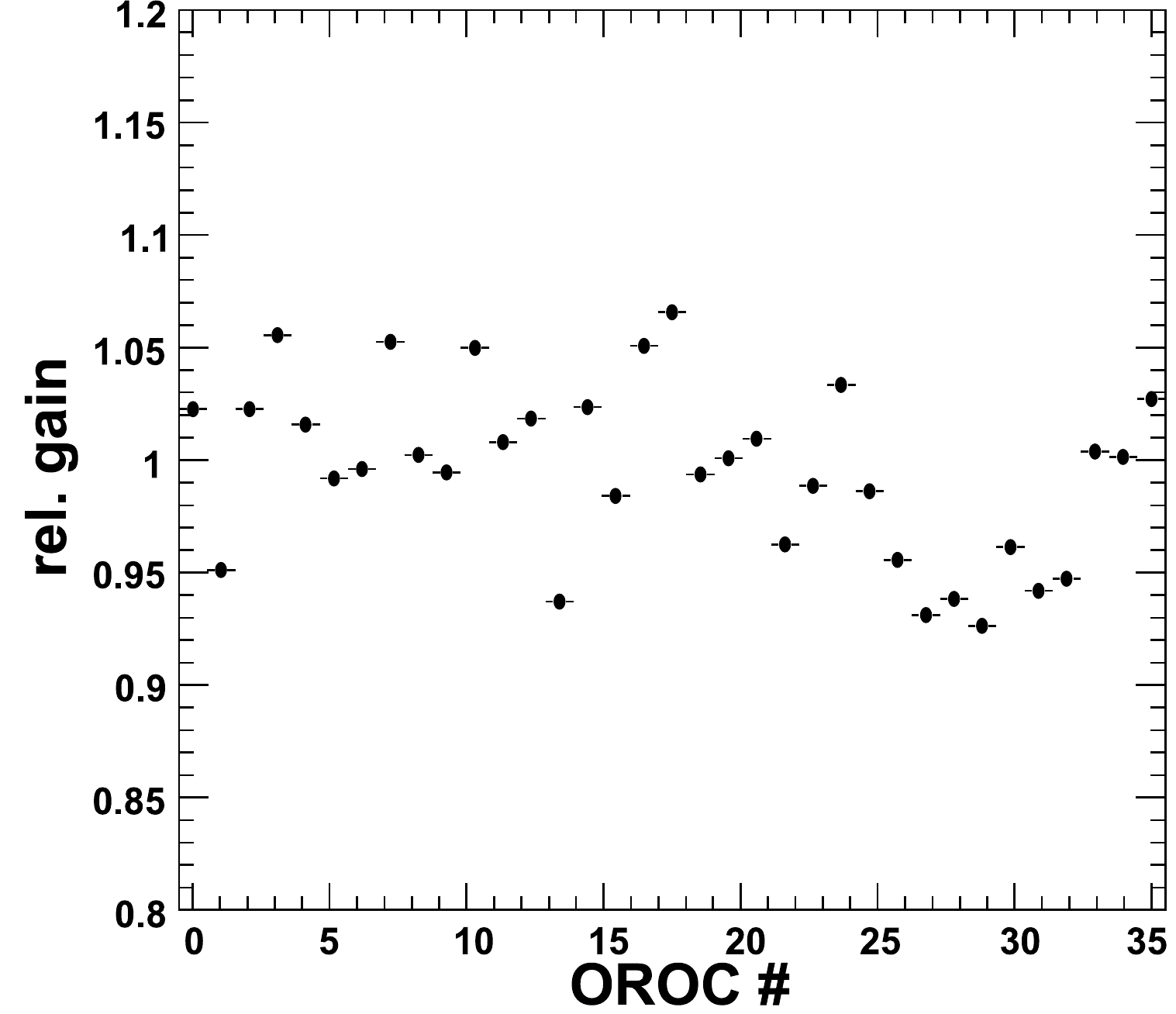}
\caption{Relative gain variations measured at fixed anode voltage for the outer
  readout chambers (OROC). The r.m.s. of this distribution is about 8\,\%.}
\label{gainpic}
\end{minipage}
\end{figure*}

\subsection{CALIBRATION WITH THE LASER SYSTEM}

A Nd:YAG laser running at 266\,nm is used to generate tracks in the TPC by sending
in 168 laser rays arranged in 4 planes along the beam direction on both sides of the
TPC\,\cite{laser}. These can be used to
align adjacent ROCs, determine electric field distortions due to space charge at
high track densities and to measure $E\times B$ effects. In addition, scattered laser
light ejects photoelectrons from the central electrode (CE). All readout pads
receive this signal at a certain characteristic time corresponding to the drift length
and velocity. This information can also be used for alignment and for drift velocity
and gain calibration. Fig. \ref{vdpic} shows the mean drift velocity calculated from
the central electrode signals. The relative resolution is $2\cdot 10^{-4}$, which is
sufficient for calibration of the data in the physics analysis. The behaviour with time
is monitored with a dedicated drift velocity monitor. The electrostatic geometry of
the ROCs was not yet tuned when the present data were taken. We expect an additional
improvement.

\subsection{GAIN AND ENERGY LOSS CALIBRATION} \label{secKr}

Due to geometric imperfections the ROC gain is not equal for all readout pads in the
TPC. For gain calibration (equalisation) three different sources are used: Krypton
data, laser (CE) data and pulser data. The Krypton method was used already by the
NA49 experiment\,\cite{na49}. Radioactive Krypton isotopes are injected in the TPC drift gas. The
charge deposit (with several peaks up to 42\,keV) is measured at random positions and
random drift times. The dynamic range of the ADC in the ALTRO chips allows to do this
at the nominal gain. Thus we are able to directly calibrate the gain of the ROCs. The
relative gain variations withing a ROC are about 7\,\% ($\sigma$). The variations
from ROC to ROC for the outer chambers (OROCs) are about 8\,\%. (r.m.s.), as shown in Fig.
\ref{gainpic}.

\begin{figure*}[th]
  \centering
  \includegraphics[width=11cm]{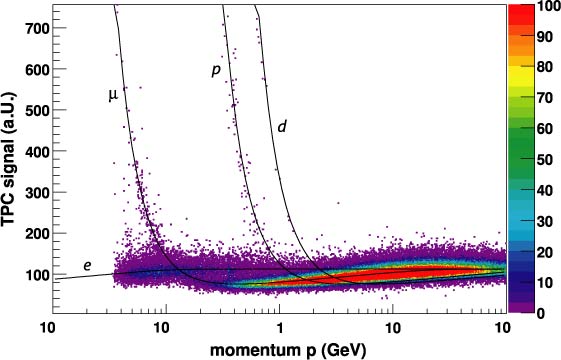}
  \caption{Energy loss distributions for various charged particles as a function of their
    momentum. The data were obtained with a cosmic ray trigger in June 2008. The lines are
    based on the Bethe-Bloch formula.}
  \label{dedxpic}
\end{figure*}

\section{PRELIMINARY RESULTS FROM THE ANALYSIS OF COSMIC TRACK DATA}

Analysing cosmic tracks in the TPC gives a first idea on how the TPC will perform
when the LHC provides collisions. We obtained first results on energy loss
(d$E$/d$x$) and momentum resolution by comparing the two parts of tracks of the same
event reconstructed in the upper and lower half of the TPC. The magnetic field was set
to $B=0.5$\,T. Fig. \ref{dedxpic} shows the energy loss as a function of momentum. The
gain calibration was done using the Krypton method described in Section
\ref{secKr}. This first analysis shows that the d$E$/d$x$-resolution is better than 6\,\%.

\begin{figure*}[th]
  \centering
  \includegraphics[width=11cm]{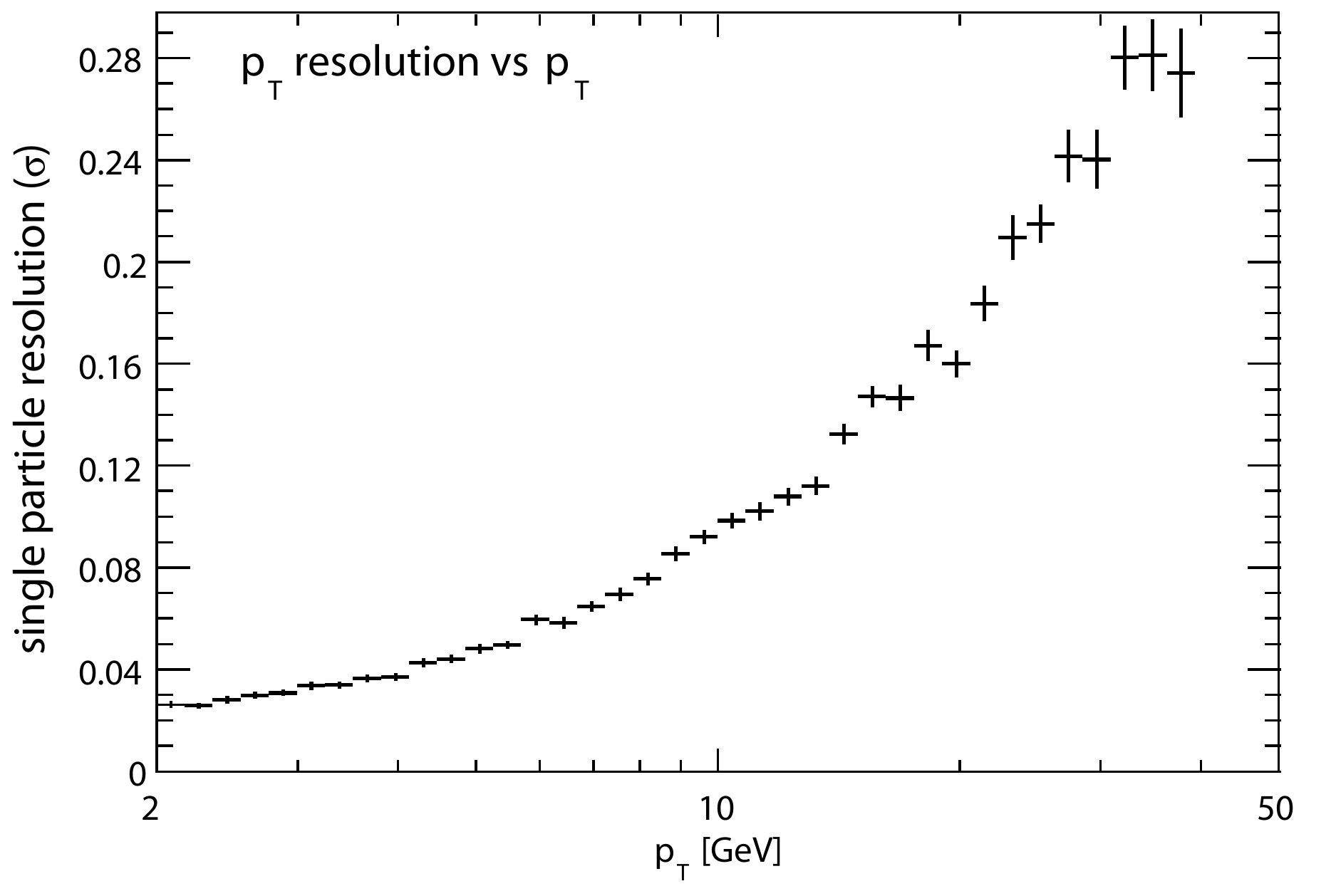}
  \caption{Momentum resolution obtained for the ALICE TPC with a cosmic ray sample. For
    details see text.}
  \label{ptpic}
\end{figure*}

Fig. \ref{ptpic} shows the momentum resolution obtained with cosmic ray tracks for the TPC.
The difference between the upper and lower part of each cosmic ray track is analyzed to
determine the momentum resolution as a function of momentum. We used data from one side of
the TPC and divided the result by $\sqrt{2}$ in order to obtain a single particle resolution.
No calibration was applied on the data; especially the $E\times B$ calibration is expected
to improve the resolution.

\section{SUMMARY}

The largest TPC ever built is at the heart of the ALICE Experiment. For the TPC readout
the ALTRO chip is used, which implements intelligent signal processing and data volume reduction
at high occupancy and rate on the Front-End-Electronics. The TPC has been fully commissioned
in the period from 2006 to summer 2008 and is now operating under nominal conditions at a
gain of about 8000. The overall system noise is close to the natural limit. Analysis of
data taken with a cosmic ray trigger gives a good indication of how the data from collisions
will look like. We currently extract the various calibration parameters from cosmic, laser and
pulser data. The TPC is ready for collisions.


\begin{thebibliography}{9}   % Use for  1-9  references
%\begin{thebibliography}{99} % Use for 10-99 references

\bibitem{alice}
  The ALICE Collaboration, K. Aamodt et al., ``The ALICE Experiment at the CERN LHC'',
  JINST 3 (2008) S08002.

\bibitem{tpctdr}
  The ALICE Collaboration, ``ALICE TPC Technical Design Report'', CERN/LHCC 2000-001,
  ALICE TDR 7, 7 January 2000.

\bibitem{chilo}
  C. Garabatos et al., Proceedings of the 10th International Vienna Conference on
  Instrumentation``The ALICE TPC'', NIM A535 (2004) 197-200.

\bibitem{musa}
  L. Musa et al., ``The ALTRO chip: a 16-channel A/D converter and digital processor for gas detectors'',
  IEEE Trans. Nucl. Sci., November 2003.

\bibitem{altro}
  B. Mota et al., ``Performance of the ALTRO chip on data aquired on an ALICE TPC prototype'',
  NIM A 535 (2004) 500-505.
  
\bibitem{laser}
  G. Renault et al., ``The Laser of the ALICE Time Projection Chamber'', Int. J. Mod. Phys. E (2007)
  2413-2418; nucl-ex/0703042.

\bibitem{na49}
  S. Wenig et al., ``Performance of the large-scale TPC system in the CERN heavy ion experiment NA49'',
  NIM A 409 (1998) 100-104.

\end{thebibliography}
\end{document}